\documentclass[a4paper]{amsart}

\RequirePackage{amsmath}
\RequirePackage{bm}
\RequirePackage{amssymb}
\RequirePackage{upref}
\RequirePackage{amsthm}
\RequirePackage{enumerate}
\RequirePackage{pb-diagram}
\RequirePackage{amsfonts}
\RequirePackage[mathscr]{eucal}
\RequirePackage{verbatim}
\RequirePackage{xr}
\RequirePackage{graphicx}
\usepackage{calc}
\usepackage{xspace}
\RequirePackage{color}

\newcommand{\cf}{cf.\@\xspace}
\newcommand{\resp}{resp.\@\xspace}

\newcommand{\al}{\alpha}
\newcommand{\bet}{\beta}
\newcommand{\ga}{\gamma}
\newcommand{\de}{\delta }
\newcommand{\e}{\epsilon}

\newcommand{\f}{\varphi}
\newcommand{\h}{\eta}

\newcommand{\ka}{\kappa}
\newcommand{\lam}{\lambda}
\newcommand{\m}{\mu}
\newcommand{\n}{\nu}
\newcommand{\om}{\omega}

\newcommand{\s}{\sigma}
\newcommand{\x}{\xi}

\newcommand{\C}{\varGamma}
\newcommand{\D}{\varDelta}

\newcommand{\Lam}{\varLambda}
\newcommand{\Om}{\varOmega}

\newcommand{\di}[1]{#1\nobreakdash-\hspace{0pt}dimensional}

\newcommand{\fv}[2]{#1\hspace{0pt}_{|_{#2}}}

\newcommand{\so}{{\mc S_0}}

\newcommand{\const}{\tup{const}}
\newcommand{\ndash}{\nobreakdash--}

\newcommand{\msp[1]}[1]{\mspace{#1mu}}

\newcommand{\R}[1][n+1]{{\protect\mathbb R}^{#1}}

\newcommand{\Cc}{{\protect\mathbb C}}

\newcommand{\N}{{\protect\mathbb N}}

\newcommand{\eR}{\stackrel{\lower1ex \hbox{\rule{6.5pt}{0.5pt}}}{\msp[3]\R[]}}
\newcommand{\eN}{\stackrel{\lower1ex \hbox{\rule{6.5pt}{0.5pt}}}{\msp[1]\N}}
\newcommand{\eO}{\stackrel{\lower1ex \hbox{\rule{6pt}{0.5pt}}}{\msc O}}
\newcommand{\mf}[1]{\mathfrak {#1}}

\DeclareMathOperator{\graph}{graph}

\DeclareMathOperator{\ad}{ad}

\DeclareMathOperator{\tr}{tr}
\DeclareMathOperator{\SO}{SO}
\DeclareMathOperator{\soc}{\mf{so}}

\newcommand\ra{\rightarrow}

\newcommand\pa{\partial}
\newcommand\pde[2]{\frac {\partial#1}{\partial#2}}
 
\newcommand{\un}{\infty}
\newcommand{\A}{\forall}

\newcommand{\set}[2]{\{\,#1\colon #2\,\}}
\newcommand{\uu}{\cup}

\newcommand{\uuu}{\bigcup}

\newcommand{\uud}{ \stackrel{\lower 1ex \hbox {.}}{\uu}}
\newcommand{\uuud}[1]{ \stackrel{\lower 1ex \hbox {.}}{\uuu_{#1}}}
\newcommand\su{\subset}

\newcommand{\sminus}[1][28]{\raise 0.#1ex\hbox{$\scriptstyle\setminus$}}

\newcommand{\wt}{\widetilde}

\newcommand{\wed}{\wedge}

\newcommand{\abs}[1]{\lvert#1\rvert}

\newcommand{\norm}[1]{\lVert#1\rVert}

\newcommand{\spd}[2]{\protect\langle #1,#2\protect\rangle}

\newcommand\ch[3]{\varGamma_{#1#2}^#3}
\newcommand\cha[3]{{\bar\varGamma}_{#1#2}^#3}

\newcommand{\riem}[4]{R_{#1#2#3#4}}
\newcommand{\riema}[4]{{\bar R}_{#1#2#3#4}}

\newcommand{\tit}{\textit}

\newcommand{\tup}{\textup}

\newcommand{\mc}{\protect\mathcal}
\newcommand{\msc}{\protect\mathscr}

\providecommand{\bysame}{\makebox[3em]{\hrulefill}\thinspace}

\newcommand{\ci}{\cite}

\newcommand{\bt}{\begin{thm}}
\newcommand{\bl}{\begin{lem}}
\newcommand{\bc}{\begin{cor}}
\newcommand{\bd}{\begin{definition}}
\newcommand{\bpp}{\begin{prop}}
\newcommand{\br}{\begin{rem}}
\newcommand{\bn}{\begin{note}}
\newcommand{\be}{\begin{ex}}
\newcommand{\bes}{\begin{exs}}
\newcommand{\bb}{\begin{example}}
\newcommand{\bbs}{\begin{examples}}
\newcommand{\ba}{\begin{axiom}}
\newcommand{\bas}{\begin{assumption}}

\newcommand{\et}{\end{thm}}
\newcommand{\el}{\end{lem}}
\newcommand{\ec}{\end{cor}}
\newcommand{\ed}{\end{definition}}
\newcommand{\epp}{\end{prop}}
\newcommand{\er}{\end{rem}}
\newcommand{\en}{\end{note}}
\newcommand{\ee}{\end{ex}}
\newcommand{\ees}{\end{exs}}
\newcommand{\eb}{\end{example}}
\newcommand{\ebs}{\end{examples}}
\newcommand{\ea}{\end{axiom}}
\newcommand{\eas}{\end{assumption}}

\newcommand{\bp}{\begin{proof}}
\newcommand{\ep}{\end{proof}}
\newcommand{\eps}{\renewcommand{\qed}{}\end{proof}}

\newcommand{\bal}{\begin{align}}

\newcommand{\bi}[1][1.]{\begin{enumerate}[\upshape #1]}
\newcommand{\bia}[1][(1)]{\begin{enumerate}[\upshape #1]}
\newcommand{\bin}[1][1]{\begin{enumerate}[\upshape\bfseries #1]}
\newcommand{\bir}[1][(i)]{\begin{enumerate}[\upshape #1]}
\newcommand{\bic}[1][(i)]{\begin{enumerate}[\upshape\hspace{2\cma}#1]}
\newcommand{\bis}[2][1.]{\begin{enumerate}[\upshape\hspace{#2\parindent}#1]}
\newcommand{\ei}{\end{enumerate}}

\newcommand\ndots{\raise 0.47ex \hbox {,}\hskip0.06em\cdots %
     \raise 0.47ex \hbox {,}\hskip0.06em} 

\newcommand{\q}{\quad}
\newcommand{\qq}{\qquad}

\newcommand{\hp}{\hphantom}

\newcommand\nd{\noindent}

\newskip\Csmallskipamount                                                
\Csmallskipamount=\smallskipamount
\newskip\Cmedskipamount
\Cmedskipamount=\medskipamount
\newskip\Cbigskipamount
\Cbigskipamount=\bigskipamount

\newcommand\cvs{\vspace\Csmallskipamount}   
\newcommand\cvm{\vspace\Cmedskipamount}

\newskip\csa
\csa=\smallskipamount

\newskip\cma
\cma=\medskipamount

\newskip\cba
\cba=\bigskipamount

\newdimen\spt
\newcommand\citem{\cvs\advance\itemno by
1{(\romannumeral\the\itemno})\hskip3pt}
\newcommand{\bitem}{\cvm\nd\advance\itemno by
1{\bf\the\itemno}\hspace{\cma}}

\newcount\itemno
\newcommand{\las}[1]{\label{S:#1}}

\newcommand{\lae}[1]{\label{E:#1}}

\newcommand{\lal}[1]{\label{L:#1}}

\newcommand{\rs}[1]{Section~\ref{S:#1}}

\newcommand{\re}[1]{\eqref{E:#1}}

\newcommand{\fre}[1]{\eqref{E:#1} on page~\tup{\pageref{E:#1}}}

\newskip\thmskip
\thmskip=\parindent

\newskip\hsk
\newenvironment{hinw}{\labelsep=0pt\begin{list}{}{\labelsep=0pt\itemindent=0pt\labelwidth=0pt\leftmargin=\parindent\rightmargin=0pt\partopsep=\cba}%
\item\it\nopagebreak\nopagebreak}%
{\end{list}}

\newcommand\bh{\begin{hinw}}
\newcommand{\eh}{\end{hinw}}

\newtheoremstyle{normal}
  {\cba}
  {\cba}
  {}
  {\thmskip}
  {\bfseries}
  {.}
  {\hsk}
  {}

\newtheoremstyle{abschnitt}
  {\cba}
  {\cba}
  {}
  {\thmskip}
  {\bfseries}
  {.}
  {\hsk}
  {}

\newtheoremstyle{italic}
  {\cba}
  {\cba}
  {\itshape}
  {\thmskip}
  {\bfseries}
  {.}
  {\hsk}
  {}

\newtheoremstyle{aufgaben}
  {\cba}
  {\cba}
  {}
  {}
  {\normalsize\bfseries}
  {.}
  {\hsk}
  {}

\newtheoremstyle{break}
  {\cba}
  {\cba}
  {\itshape}
  {}
  {\bfseries}
  {.}
  {\newline}
  {}

\swapnumbers
\theoremstyle{italic}
\newtheorem{thm}[subsection]{Theorem}
\newtheorem{lem}[subsection]{Lemma}
\newtheorem{prop}[subsection]{Proposition}
\newtheorem{cor}[subsection]{Corollary}

\theoremstyle{normal}
\newtheorem{rem}[subsection]{Remark}
\newtheorem{definition}[subsection]{Definition}
\newtheorem{example}[subsection]{Example}
\newtheorem{examples}[subsection]{Examples}
\newtheorem{ex}[subsection]{Exercise}
\newtheorem{note}[subsection]{}
\newtheorem{axiom}[subsection]{Axiom}
\newtheorem{assumption}[subsection]{Assumption}

\theoremstyle{aufgaben}
\newtheorem{exs}[subsection]{Exercises}

\swapnumbers

\numberwithin{equation}{section}
\numberwithin{figure}{section}

\newenvironment{textequation}[1][0.8]
{\begin{equation}
\begin{aligned}
\begin{minipage}{#1\linewidth}}
{\end{minipage}
\end{aligned}
\end{equation}
\ignorespacesafterend}

\newcommand{\btext}{\begin{textequation}}
\newcommand{\etext}{\end{textequation}}

\def\hinweis{\@startsection{subsection}{2}%
 \z@{0.7\linespacing\@plus 0.5\linespacing}{0.7\linespacing}%
{\normalfont\itshape\indent}}

\usepackage[german,english]{babel}
\usepackage{graphicx}
\RequirePackage{amsmath}
\RequirePackage{bm}
\RequirePackage{amssymb}
\RequirePackage{upref}
\RequirePackage{amsthm}
\RequirePackage{enumerate}
\RequirePackage{pb-diagram}
\RequirePackage{amsfonts}
\RequirePackage[mathscr]{eucal}
\RequirePackage{verbatim}
\RequirePackage{xr}
\RequirePackage{graphicx}
\usepackage{calc}
\usepackage{xspace}

\makeatletter
\RequirePackage{color}
\newcommand{\ann}[1]{\renewcommand{\@makefnmark}{\mbox{$^{\color{red}{\@thefnmark}}$}}%
\footnote {#1}}
\makeatother

\RequirePackage{upref}
\RequirePackage{amsthm}
\RequirePackage{enumerate}
\usepackage[mathscr]{eucal}

\usepackage{xr-hyper}

\usepackage{calc}

\newlength{\oddsidemarginlength}
\newlength{\topmarginlength}

\newcounter{numberoflines}
\newcounter{tempcc}
\oddsidemargin=\oddsidemarginlength
\evensidemargin=\oddsidemargin
\topmargin=\topmarginlength
\begin{document}
\flushbottom

\title[Quantum cosmological models]{Quantum cosmological Friedman models with a massive Yang-Mills field}

\author{Claus Gerhardt}
\address{Ruprecht-Karls-Universit\"at, Institut f\"ur Angewandte Mathematik,
Im Neuenheimer Feld 294, 69120 Heidelberg, Germany}
\email{gerhardt@math.uni-heidelberg.de}
\urladdr{http://www.math.uni-heidelberg.de/studinfo/gerhardt/}
\thanks{This work has been supported by the DFG}

%
\date{\today}
%


\begin{abstract}
We prove the existence of a spectral resolution of the Wheeler-DeWitt equation when the matter field is provided by a massive Yang-Mills field. The resolution is achieved by first solving the free eigenvalue problem for the gravitational field and then the constrained eigenvalue problem for the Yang-Mills field. In the latter case the mass of the Yang-Mills field assumes the role of the eigenvalue.
\end{abstract}

\maketitle

\tableofcontents

\setcounter{section}{0}
\section{Introduction}

In \cite{cg:qfriedman} we proved the existence of a complete spectral resolution of the Wheeler-DeWitt equation when the matter Lagrangian corresponds to that of scalar field. A Yang-Mills Lagrangian is of course especially interesting and a corresponding quantum cosmological model providing a spectral resolution of the Wheeler-DeWitt equation would be most desirable. 

The method developed in our previous paper for solving the Wheeler-DeWitt equation comprises three steps: First, the Hamilton operators corresponding to the gravitational field and the matter field, respectively, have to be separated; second, for one of the operators a complete set of eigenfunctions has to be found, i.e., a \tit{free} spectral resolution has to be proved without any constraints; third, for the remaining Hamilton operator then a \tit{constrained} spectral resolution has to be found by looking at the Wheeler-DeWitt equation as an \tit{implicit} eigenvalue problem. 

Solving the implicit eigenvalue problem requires that a lower order term of the operator carries a scalar factor that can serve as an eigenvalue. If the Hamilton operator of the gravitational field is used in this step, then the cosmological constant can play this role. However, if the matter Hamiltonian is involved then the corresponding matter Lagrangian must contain a variable scalar factor usually representing the mass of the field. 

A satisfactory quantum cosmological model would both have to be a solution as well as to be unique once the spatial topology of the underlying classical Friedman model has been fixed. By uniqueness we also mean that one has no choice which operator to use in the second step, since the resulting models would be different generally.

In case of a scalar field always the matter Hamiltonian has to be chosen in the second step, and  the so-called unbounded model\footnote{The unbounded model is characterized by the fact that the scale factor ranges from zero to infinity, and we note that uniqueness of the model necessarily requires an unbounded range of the scale factor.} in \cite{cg:qfriedman} satisfies the further requirement of uniqueness. When considering Yang-Mills fields  an unbounded model is only achievable, if the gravitational Hamilton operator is used for the free eigenvalue problem. Hence, the Yang-Mills Lagrangian must necessarily involve a mass term for otherwise the constrained eigenvalue problem cannot be solved. 

The eigenvalues of the final solution are the implicit eigenvalues of the Hamilton operator of the Yang-Mills field which correspond to the mass of the Yang-Mills field. 

We shall prove the existence of two unbounded solutions corresponding to the topologies of the spatial sections, which are given by either $\R[3]$ or $S^3$. 

Classically, we look at  spatially homogeneous spacetimes $N=N^{4}$ where the Lorentzian metrics are of the form
\begin{equation}\lae{1.1}
d\bar s^2=-w^2 dt^2+r^2\s_{ij}(x)dx^idx^j;
\end{equation}
here $(\s_{ij})$ is the metric of a simply connected space of constant curvature $\so$ with curvature $\tilde\ka\in\{0,1\}$, and $r$, $w$ are positive functions depending only on $t$, and the Einstein equations are the Euler-Lagrange equation of   the functional \begin{equation}\lae{1.2.6}
J=\int_N(\bar R-2\Lam)+\al_MJ_M,
\end{equation}
where $\bar R$ is the scalar curvature, $\Lam$ a cosmological constant, $\al_M$ a positive coupling constant, and $J_M$ a functional representing matter, which will be the curvature functional of a massive Yang-Mills field with values in $\mf {so}(3)\otimes T^{0,1}(N)$.

The Yang-Mills Lagrangian then consists of the curvature squared and a mass term
\begin{equation}
L_M=\tfrac14\tr(F_{\mu\lam} F^{\mu\lam})+\tfrac \mu2 m(A),
\end{equation}
where $A$ represents the connection and $\mu$ is a real parameter which will be the mass.

When dealing with Yang-Mills connections one usually assumes that there exists a globally defined reference connection $\bar A=(\bar A_\mu)$, then the difference with another connection $A=(A_\mu)$ is a tensor
\begin{equation}
A_\mu-\bar A_\mu=f_aA^a_\mu,
\end{equation}
where $(f_a)$ is a basis of $\ad(\mf g)$ and $A^a_\mu$ a $\mf g$-valued 1-form. The Lie algebra $\mf g$ is the fiber of the adjoint bundle.

Then a natural mass density would be
\begin{equation}
m(A)=\ga_{ab}A^a_\mu A^b_\lam \bar g^{\mu\lam},
\end{equation}
where $\ga_{ab}$ is the Cartan-Killing metric in $\mf g$ and $\bar g_{\mu\lam}$ the metric in $N$.

$m(A)$ is obviously fully covariant and the Euler-Lagrange equation of the functional
\begin{equation}
\int_\Om \tfrac14 \tr(F_{\mu\lam} F^{\mu\lam})+\tfrac\mu2\int_\Om m(A),
\end{equation}
where $\Om\su N$ is open and relatively compact, is
\begin{equation}\lae{1.70}
F^{a\mu}_{\hp{a\mu}\lam;\mu}+\mu A^a_\lam=0
\end{equation}
in $\Om$.

In the Friedman cosmological case we look at very special Yang-Mills fields.
Let $\so$ be one of the homogeneous Riemannian spaces $\R[3]$ or $ S^3$ corresponding to the Lie groups of Bianchi type I or IX, respectively, which act transitively on these spaces. An old result of Bianchi states that there exists a set of left invariant 1-forms $\om^a$, $a=1,2,3$, on $\so$ such that
\begin{equation}\lae{1.3}
\s_{ij}=\de_{ab}\om^a_i\om^b_j\q\wed\q \s^{ij}\om^a_i\om^b_j=\de^{ab}
\end{equation}
and
\begin{equation}\lae{1.4}
d\om^a=\tfrac12c^a_{bc}\om^b\om^c,
\end{equation}
where $c^a_{bc}$ are the structure constants of the corresponding group, i.e., 
\begin{equation}
c^a_{bc}=
\begin{cases}
0,& \text{Bianchi type I}\\
\e^a_{bc}, &\text{the Levi-Civit\`a symbol}, \text{Bianchi type IX}\\
\end{cases}
\end{equation}
\cf \cite[Appendix]{taub} or \cite[p. 110]{ryan:book}, see also \cite{helgason}. An existence  proof for the 1-forms satisfying \re{1.3} and \re{1.4} is given in \cite[Chap. 77 and Appendix 29]{eisenhart}.  We also note that $\SO(3)$ is of type IX.

We extend these 1-forms to the spacetime $N$ by setting 
\begin{equation}
\tilde\om^a=(\tilde\om^a_\mu)=(0,\om^a_i)
\end{equation}
and define a tensor field
\begin{equation}
A^a_\mu=
\f \tilde\om^a_\mu\in \soc(3)\otimes T^{0,1}(N),
\end{equation}
where $\f=\f(t)$ is a real function, which will generate a connection in the adjoint bundle.

It is well-known that an $\soc(3)$-bundle over $S^3$ is trivial, i.e., there exists a smooth triple $(G_a)$ of sections forming a basis of $\mf \soc(3)$. Since $\R[3]$ can be viewed as an open subset of $S^3$, the same is valid for an $\soc(3)$-bundle over $\R[3]$. In trivial bundles there exists a pure gauge connection. These results remain valid, if the base space is a product $I\times \so$, $I$ an interval, like in the cosmological case.

Hence, we shall use as reference connection the globally defined pure gauge connection $\bar A=(\bar A_\mu)$. 

We then consider the functional
\begin{equation}
J=\int_\Om (\bar R-2\Lam)+\al_M\int_\Om\{-\tfrac14 \ga_{ab}F^a_{\mu\lam}F^{b\mu\lam}+\tfrac{\mu}2 \ga_{ab}A^a_\mu A^b_\lam \bar g^{\mu\lam}\chi_0^{-\frac23}\},
\end{equation}
where $\al_M$ is a positive coupling constant, and $\Om\su N$ is open and such that
\begin{equation}
\Om=I\times\tilde\Om,
\end{equation}
where $I=(a,b)$ is a bounded interval and $\tilde\Om\su\so$ an arbitrary open set of measure one with respect to the standard metric of $\so$.

$\chi_0=\chi_0(x,\bar g_{\al\bet})$ is a function defined in an open set of $T^{0,2}(N)$ with the properties that it will preserve the perfect fluid structure of the energy momentum tensor and will lead to the right powers of the scale factor when switching from the Lagrangian view to the Hamiltonian view. If $\bar g_{\al\bet}$ is a product metric then $\chi_0=1$.

The reasons for the introduction of $\chi_0$ are explained in \rs 4.

A canonical quantization of this functional, without the mass term, has already been treated in \cite{cavaglia}, however, no spectral resolution of the Wheeler-DeWitt equation  has been achieved; see also \cite{henneaux:830}.

We shall prove:
\bt
Assuming that
\begin{equation}
\Lam<0\qq\text{or}\qq \Lam\le 0\q\wed\q \tilde\ka=1,
\end{equation}
then there exists a self-adjoint operator $H$ in a Hilbert space $\mc H_0$, which can be looked at as a dense subspace of $L^2(\R[]_+\times\R[],\Cc)$, with a pure point spectrum consisting of countably many eigenvalues such that the eigenfunctions, and the elements of the vector space $W$ generated by them, are solutions of the Wheeler-DeWitt equation.

The solutions of the corresponding Schr\"odinger equation, with initial values $\psi_0\in W,$ provide a dynamical development of the quantum model.
\et

The paper is organized as follows:

The quantization of the Lagrangian is derived in \rs 4, while the spectral resolution is achieved in \rs 5 and \rs 6.

\section{Notations and definitions}\las{01}

The main objective of this section is to state the equations of Gau{\ss}, Codazzi,
and Weingarten for spacelike hypersurfaces $M$ in a \di {(n+1)} Lorentzian
manifold
$N$.  Geometric quantities in $N$ will be denoted by
$(\bar g_{ \al \bet}),(\riema  \al \bet \ga \de)$, etc., and those in $M$ by $(g_{ij}), 
(\riem ijkl)$, etc.. Greek indices range from $0$ to $n$ and Latin from $1$ to $n$;
the summation convention is always used. Generic coordinate systems in $N$ resp.
$M$ will be denoted by $(x^ \al)$ \resp $(\x^i)$. Covariant differentiation will
simply be indicated by indices, only in case of possible ambiguity they will be
preceded by a semicolon, i.e., for a function $u$ in $N$, $(u_ \al)$ will be the
gradient and
$(u_{ \al \bet})$ the Hessian, but e.g., the covariant derivative of the curvature
tensor will be abbreviated by $\riema  \al \bet \ga{ \de;\e}$. We also point out that
\begin{equation}
\riema  \al \bet \ga{ \de;i}=\riema  \al \bet \ga{ \de;\e}x_i^\e
\end{equation}
with obvious generalizations to other quantities.

Let $M$ be a \tit{spacelike} hypersurface, i.e., the induced metric is Riemannian,
with a differentiable normal $\n$ which is timelike.

In local coordinates, $(x^ \al)$ and $(\x^i)$, the geometric quantities of the
spacelike hypersurface $M$ are connected through the following equations
\begin{equation}\lae{01.2}
x_{ij}^ \al= h_{ij}\n^ \al
\end{equation}
the so-called \tit{Gau{\ss} formula}. Here, and also in the sequel, a covariant
derivative is always a \tit{full} tensor, i.e.

\begin{equation}
x_{ij}^ \al=x_{,ij}^ \al-\ch ijk x_k^ \al+ \cha  \bet \ga \al x_i^ \bet x_j^ \ga.
\end{equation}
The comma indicates ordinary partial derivatives.

In this implicit definition the \tit{second fundamental form} $(h_{ij})$ is taken
with respect to $\n$.

The second equation is the \tit{Weingarten equation}
\begin{equation}
\n_i^ \al=h_i^k x_k^ \al,
\end{equation}
where we remember that $\n_i^ \al$ is a full tensor.

Finally, we have the \tit{Codazzi equation}
\begin{equation}
h_{ij;k}-h_{ik;j}=\riema \al \bet \ga \de\n^ \al x_i^ \bet x_j^ \ga x_k^ \de
\end{equation}
and the \tit{Gau{\ss} equation}
\begin{equation}
\riem ijkl=- \{h_{ik}h_{jl}-h_{il}h_{jk}\} + \riema  \al \bet\ga \de x_i^ \al x_j^ \bet
x_k^ \ga x_l^ \de.
\end{equation}

Now, let us assume that $N$ is a globally hyperbolic Lorentzian manifold with a
\tit{compact} Cauchy surface. 
$N$ is then a topological product $I\times \mc S_0$, where $I$ is an open interval,
$\mc S_0$ is a compact Riemannian manifold, and there exists a Gaussian coordinate
system
$(x^ \al)$, such that the metric in $N$ has the form 
\begin{equation}\lae{01.7}
d\bar s_N^2=e^{2\psi}\{-{dx^0}^2+\s_{ij}(x^0,x)dx^idx^j\},
\end{equation}
where $\s_{ij}$ is a Riemannian metric, $\psi$ a function on $N$, and $x$ an
abbreviation for the spacelike components $(x^i)$. 
We also assume that
the coordinate system is \tit{future oriented}, i.e., the time coordinate $x^0$
increases on future directed curves. Hence, the \tit{contravariant} timelike
vector $(\x^ \al)=(1,0,\dotsc,0)$ is future directed as is its \tit{covariant} version
$(\x_ \al)=e^{2\psi}(-1,0,\dotsc,0)$.

Let $M=\graph \fv u\so$ be a spacelike hypersurface
\begin{equation}
M=\set{(x^0,x)}{x^0=u(x),\,x\in\mc S_0},
\end{equation}
then the induced metric has the form
\begin{equation}
g_{ij}=e^{2\psi}\{-u_iu_j+\s_{ij}\}
\end{equation}
where $\s_{ij}$ is evaluated at $(u,x)$, and its inverse $(g^{ij})=(g_{ij})^{-1}$ can
be expressed as
\begin{equation}\lae{01.10}
g^{ij}=e^{-2\psi}\{\s^{ij}+\frac{u^i}{v}\frac{u^j}{v}\},
\end{equation}
where $(\s^{ij})=(\s_{ij})^{-1}$ and
\begin{equation}\lae{01.11}
\begin{aligned}
u^i&=\s^{ij}u_j\\
v^2&=1-\s^{ij}u_iu_j\equiv 1-\abs{Du}^2.
\end{aligned}
\end{equation}
Hence, $\graph u$ is spacelike if and only if $\abs{Du}<1$.

The covariant form of a normal vector of a graph looks like
\begin{equation}
(\n_ \al)=\pm v^{-1}e^{\psi}(1, -u_i).
\end{equation}
and the contravariant version is
\begin{equation}
(\n^ \al)=\mp v^{-1}e^{-\psi}(1, u^i).
\end{equation}
Thus, we have
\br Let $M$ be spacelike graph in a future oriented coordinate system. Then the
contravariant future directed normal vector has the form
\begin{equation}
(\n^ \al)=v^{-1}e^{-\psi}(1, u^i)
\end{equation}
and the past directed
\begin{equation}\lae{01.15}
(\n^ \al)=-v^{-1}e^{-\psi}(1, u^i).
\end{equation}
\er

In the Gau{\ss} formula \re{01.2} we are free to choose the future or past directed
normal, but we stipulate that we always use the past directed normal for reasons
that we have explained in \ci[Section 2]{cg:indiana}.

Look at the component $ \al=0$ in \re{01.2} and obtain in view of \re{01.15}

\begin{equation}\lae{01.16}
e^{-\psi}v^{-1}h_{ij}=-u_{ij}- \cha 000\mspace{1mu}u_iu_j- \cha 0j0
\mspace{1mu}u_i- \cha 0i0\mspace{1mu}u_j- \cha ij0.
\end{equation}
Here, the covariant derivatives are taken with respect to the induced metric of
$M$, and
\begin{equation}
-\cha ij0=e^{-\psi}\bar h_{ij},
\end{equation}
where $(\bar h_{ij})$ is the second fundamental form of the hypersurfaces
$\{x^0=\const\}$.

An easy calculation shows
\begin{equation}
\bar h_{ij}e^{-\psi}=-\tfrac{1}{2}\dot\s_{ij} -\dot\psi\s_{ij},
\end{equation}
where the dot indicates differentiation with respect to $x^0$.

\section{The quantization of the Lagrangian}\las 4  

Let us first derive the matter Lagrangian $L_M$ for the problem. It has to have three properties: first, it is a function on the underlying tensor spaces, i.e., it is invariant under gauge and coordinate  transformations, second, in the cosmological case, when the Euler-Lagrange equations are merely the Friedman equations and not the full Einstein equations, the corresponding energy momentum tensor has to be that of a perfect fluid, for otherwise the Friedman equation is not equivalent to the the Einstein equations, and third, the resulting Wheeler-DeWitt equation should be solvable, i.e., a spectral resolution should be possible, at least in principle.

The last point needs some elaboration. The ans\"atze for the matter Lagrangian in QFT with the Minkowski space as underlying spacetime, when applied to curved spacetimes, very often lead to Wheeler-DeWitt equations for which a spectral resolution is not possible, at least not by the method outlined in the introduction, since the lower order terms of the matter Hamiltonian, containing the potential, are equipped with a different power of the scale factor than the leading term, making a separation of the Hamiltonian of the gravitational field from that of the matter field impossible. This happened in \cite{cg:qfriedman} where we therefore had to consider a scalar field \tit{without} potential.

In the case of a massive Yang-Mills field we are confronted with the same dilemma, or an even greater dilemma, since without the mass term the Wheeler-DeWitt equation is not solvable either. Hence, we modify the lower order terms in the matter Lagrangian slightly such that they lead to the right powers of the scale factor, when considered in curved spacetimes, but agree with corresponding terms of the Lagrangian in QFT when the spacetime is Minkowskian.

The leading term of the matter Lagrangian is the squared curvature of the connection
\begin{equation}
-\tfrac14 R_{ab\mu\lam}R^{ab\mu\lam}=-\tfrac14 \ga_{ab}F^a_{\mu\lam}F^{b\mu\lam}.
\end{equation}

The lower order term would ideally be of the form
\begin{equation}\lae{4.2}
\tfrac{\mu}2 m(A),
\end{equation}
where $m(A)$ is the matter density and $\mu\in\R[]$ representing the mass. However, using \re{4.2} unaltered will lead to the wrong powers of the scale factor.

Therefore, we instead use
\begin{equation}
\tfrac{\mu}2 m(A)\chi_0^p,
\end{equation}
where $p\in\R[]$ is an appropriate exponent and $\chi_0$ a positive function defined in the space of all spacetime metrics as follows:

\bl
Let $N^{n+1}=(N,\bar g_{\al\bet})$ be a globally hyperbolic spacetime with Cauchy hypersurface $\so$ such that topologically $N=I\times \so$ and let $\s_{ij}$ be a fixed Riemannian metric on $\so$. Choose a coordinate system $(x^\al)$ such that
\begin{equation}\lae{4.4}
d\bar s^2=-w^2 (dx^0)^2+\bar g_{ij}dx^idx^j
\end{equation}
and define a metric $\tilde g_{\al\bet}$ through the definition
\begin{equation}\lae{4.5}
d\tilde s^2=-w^2 (dx^0)^2+\s_{ij}dx^idx^j
\end{equation}
in that particular coordinate system. Then $\tilde g_{\al\bet}$ is a Lorentz metric in any coordinate system and
\begin{equation}
\chi_0=\frac{\det(\bar g_{\al\bet})}{\det(\tilde g_{\al\bet})}
\end{equation}
is a function on $N$, or more generally, a function on
\begin{equation}
\C(N)=\set{(\bar g_{\al\bet})\in T^{0,2}(N)}{ \bar g_{\al\bet}\; Lorentzian, \, \so \;\text{C. hypersurface}},
\end{equation}
such that
\begin{equation}
\pde{\chi_0}{\bar g^{\al\bet}}=
\begin{cases}
-\chi_0\bar g_{ij}, &(\al,\bet)=(i,j),\\
0,& \text{else},
\end{cases}
\end{equation}
if the derivative is evaluated at a metric satisfying \re{4.4} in an appropriate coordinate system. In this case $\chi_0$ can also be expressed as
\begin{equation}
\chi_0=\frac{\det(\bar g_{ij})}{\det(\s_{ij})}.
\end{equation}
\el

The proof is straight-forward.

Thus, a factor of the form $\chi_0^p$ will give us any desired power of the scale factor and will also preserve the perfect fluid structure of the energy momentum tensor.

In our particular case we shall choose
\begin{equation}
p=-\tfrac23.
\end{equation}

Hence the ansatz for the matter Lagrangian is
\begin{equation}
L_M=-\tfrac14 \ga_{ab}F^a_{\mu\lam}F^{b\mu\lam}+\tfrac{\mu}2 \ga_{ab}A^a_\mu A^b_\lam \bar g^{\mu\lam}\chi_0^{-\frac23},
\end{equation}
where of course $\chi_0$ is defined by setting the metric $\s_{ij}$ in \re{4.5} equal to the standard metric of the respective $\so$.

Next, we have to verify that the energy momentum tensor has the structure of a perfect fluid, which can be easily checked.

Thus, we consider the functional
\begin{equation}
J=\int_\Om (\bar R-2\Lam)+\al_M\int_\Om\{-\tfrac14 \ga_{ab}F^a_{\mu\lam}F^{b\mu\lam}+\tfrac{\mu}2 \ga_{ab}A^a_\mu A^b_\lam \bar g^{\mu\lam}\chi_0^{-\frac23}\},
\end{equation}
where $\al_M$ is a positive coupling constant, and $\Om\su N$ is open and such that
\begin{equation}
\Om=I\times\tilde\Om,
\end{equation}
where $I=(a,b)$ is a bounded interval and $\tilde\Om\su\so$ an arbitrary open set of measure  one with respect to the standard metric of $\so$.

We use the action principle that for an arbitrary $\Om$ as above a solution $(A,\bar g)$ should be a stationary point of the functional with respect to compact variations. This principle requires no additional surface terms in the functional. The resulting Euler-Lagrange equations will be the Einstein equations as well as the Yang-Mills equation
\begin{equation}
F^{a\mu}_{\hp{a\mu}\lam;\mu}+\mu A^a_\lam\chi_0^{-\frac23}=0,
\end{equation}
\cf \fre{1.70}, which now incorporates the additional term $\chi_0^{-\frac23}$ being equal to $1$, if the metric of $N$ is a product metric, or even more generally, of the form \re{4.5}. 

In case of the special tensor field $(A^a_\mu)$ and the special spacetime metric, the matter Lagrangian can be expressed as
\begin{equation}
L_M=\dot\f^2w^{-2}e^{-2f}-3(\tilde\ka\f-\f^2)^2e^{-4f}+6\mu \f^2e^{-4f},
\end{equation}
where $\tilde\ka=0,1$ refers to the different choices for $\so$ and accidentally agrees with their respective curvatures, and where we used that the metric is of the form \fre{1.1}. The scale factor $r$ has been expressed in the form 
\begin{equation}
r=e^{f}.
\end{equation}

Arguing as in \cite[Section 3]{cg:qfriedman}, we conclude that the functional is equal to
\begin{equation}
\begin{aligned}
J=&\int_a^b\{6\tilde\ka e^fw-6\abs{f'}^2e^{3f}w^{-1}-2\Lam e^{3f}w\}\\
&+\al_M\int_a^b\{\dot\f^2e^fw^{-1}-3(\tilde\ka\f+\f^2)^2e^{-f}w+6\mu \f^2e^{-f}w\},
\end{aligned}
\end{equation}
or equivalently, after dividing by $6$, but calling the resulting functional still $J$,
\begin{equation}
\begin{aligned}
J=&\int_a^b\{\tilde\ka e^fw-\abs{f'}^2e^{3f}w^{-1}-\frac13\Lam e^{3f}w\}\\
&+\frac{\al_M}6\int_a^b\{\dot\f^2e^fw^{-1}-3(\tilde\ka\f+\f^2)^2e^{-f}w+6\mu \f^2e^{-f}w\}.
\end{aligned}
\end{equation}
Here, a dot or prime indicates differentiation with respect to the time $x^0$. 

Thus, our functional depends on the variables $(f,\f,w)$.

Before we apply the Legendre transformation, let us express the quadratic terms involving the derivatives with the help of a common metric. 

For $0\le a,b\le 1$ define
\begin{equation}
(y^a)=(y^0,y^1)=(f,\f),
\end{equation}
\begin{equation}
\big(\tilde G_{ab}\big)=\begin{pmatrix} 
-e^{3f}&0\\[\cma]
0&\bar\al_M e^f
\end{pmatrix},
\end{equation}
\begin{equation}
V=2\al_M(\tilde\ka\f+\f^2)^2,
\end{equation}
\begin{equation}
\bar\al_M=\frac{2\al_M}3,
\end{equation}
and
\begin{equation}
\bar\Lam=\frac4{3}\Lam\q\wed\q \bar\mu=4\al_M\mu.
\end{equation}
Then $J$ can be expressed as
\begin{equation}
\begin{aligned}
\int_a^bL=\int_a^bw\{\tilde G_{ab}\dot y^a\dot y^bw^{-2}-\tfrac14 \bar\Lam e^{3f}+\tilde\ka e^f-\tfrac14 Ve^{-f}w+ \tfrac14\bar\mu \f^2e^{-f}\}.
\end{aligned}
\end{equation}

Applying now the Legendre transformation we obtain the Hamiltonian $\tilde H$
\begin{equation}
\begin{aligned}
\tilde H=\tilde H(w,y^a,p_a)&=\dot y^a\pde L{\dot y^a}-L\\
&=\{\wt G_{ab} \dot y^a\dot y^bw^{-2} + \tfrac14 V e^{-f}+ \tfrac14\bar\Lam e^{3f}-\tilde\ka e^{f}-\tfrac14\bar\mu\f^2e^{-f}\}w\\
&=\{\tfrac14\wt G^{ab}p_ap_b+ \tfrac14 V e^{-f}+ \tfrac14\bar\Lam e^{3f}-\tilde\ka e^{f}-\tfrac14\bar\mu\f^2e^{-f}\}w\\
&\equiv Hw,
\end{aligned}
\end{equation}
and the Hamiltonian constraint requires  
\begin{equation}
H(y^a,p_a)=0.
\end{equation}

Canonical quantization stipulates to replace the momenta $p_a$ by
\begin{equation}
p_a=\frac{\hbar}i\frac{\pa}{\pa y^a}.
\end{equation}

Hence, using the convention $\hbar=1$, we conclude that the Hamilton operator $H$ is equal to
\begin{equation}
H=-\tfrac14\tilde\D+ \tfrac14 Ve^{-f}+ \tfrac14\bar\Lam e^{3f}-\tilde\ka e^{f}-\tfrac14\bar\mu\f^2e^{-f}.
\end{equation}

Note that the metric $\wt G_{ab}$ is a Lorentz metric, i.e., $H$ is hyperbolic.

Let $\psi=\psi(y)$ be a smooth function then
\begin{equation}
\tilde\D\psi=\frac1{\sqrt\abs{\wt G}}\frac\pa{\pa y^a}\bigg(\sqrt{\abs{\wt G}}\wt G^{ab}\psi_b\bigg).
\end{equation}

Now
\begin{equation}
\abs{\wt G}=\bar\al_Me^{4f}
\end{equation}
and hence
\begin{equation}\lae{4.31}
-\tilde\D\psi=e^{-2f}\frac\pa{\pa y^0}\big(e^{-f}\pde\psi{y^0}\big)-e^{-f}\bar\al_M^{-1}\frac{\pa^2\psi}{(\pa y^1)^2}.
\end{equation}

Defining a new variable
\begin{equation}
r=e^{f}=e^{y^0},
\end{equation}
setting $y=y^1$, $c_1=\bar\al_M^{-1}$ and stipulating that a dot indicates a differentiation with respect to $r$ and a prime with respect to $y$, we conclude that the Wheeler-DeWitt equation equals
\begin{equation}
r^{-1}\tfrac14 \Ddot\psi-r^{-1}\tfrac14 c_1 \psi''+ \tfrac14 Vr^{-1}\psi+ \tfrac14\bar\Lam r^3\psi-\tilde\ka r\psi-\tfrac14\bar\mu y^2r^{-1}\psi=0,
\end{equation}
or equivalently, after multiplying with $4r>0$,
\begin{equation}
\Ddot\psi-c_1\psi''+V\psi+\bar\Lam  r^4\psi-4\tilde\ka r^2\psi-\bar\mu y^2\psi=0.
\end{equation}

Thus, we have proved:
\bt
The Wheeler-DeWitt equation has the form
\begin{equation}\lae{4.35}
H_2\psi-H_1\psi-\bar\mu y^2\psi=0,
\end{equation}
where
\begin{equation}
H_1\psi=-\Ddot\psi-\bar\Lam r^4\psi +4\tilde\ka r^2\psi
\end{equation}
and
\begin{equation}
H_2\psi=-c_1\psi''+V\psi.
\end{equation}
The wave function $\psi$ is defined in a suitable subspace of $L^2(\R[]_+\times\R[],\Cc)$.
\et

\section{The eigenvalue problems}\las{5}
The Hamiltonian in the Wheeler-DeWitt equation \re{4.35} is already separated, hence, a separation of variables is possible
\begin{equation}
\psi(r,y)=u(r)\h(y),\qq (r,y)\in\R[]_+\times\R[].
\end{equation}

We now first solve the free eigenvalue problem for $H_1$,
\begin{equation}\lae{5.2}
H_1u=\lam u,
\end{equation}
where, for simplicity, we assume without loss of generality, $u$ to be real valued.

To find a complete set of eigenfunctions, the energy form of the operator
\begin{equation}
\spd{H_1 u}u=\int_{\R[]_+}\{\dot u^2-\bar\Lam r^4u^2+4\tilde\ka r^2u^2\},
\end{equation}
must be so strong that the quadratic form
\begin{equation}
K(u)=\int_{\R[]_+}u^2
\end{equation}
is compact relative to this form in a suitable Sobolev space.

Assuming either
\begin{equation}
\bar\Lam <0\qq\text{or}\qq\bar\Lam\le 0\q\wed\q\tilde\ka=1
\end{equation}
it is evident that  this requirement is satisfied in the Hilbert space
\begin{equation}
\mc H_1=\set{u\in L^2({\R[]_+})}{\norm u_1^2=\int_{\R[]_+}(\dot u^2+r^qu^2)<\un},
\end{equation}
where 
\begin{equation}\lae{5.70}
q=
\begin{cases}
4,&\Lam<0,\\
2,&\Lam=0.
\end{cases}
\end{equation}

The eigenvalue problem \re{5.2} can then be phrased as: Find pairs 
\begin{equation}
(\lam,u)\in \R[]\times \mc H_1
\end{equation} 
such that
\begin{equation}
\spd{H_1 u}v=\lam \spd uv=\lam K(u,v)\qq\A\, v\in \mc H_1,
\end{equation}
both scalar products can be viewed as the scalar product in $L^2(\R[]_+)$, where at the left-hand side we applied partial integration.

A proof that the quadratic form $K$ is compact in $\mc H_1$ can be deduced from the proof of a similar result in \cite[Lemma 6.8]{cg:qfriedman}.

From a general existence theorem for eigenvalue problems of this kind, \cf, e.g., \cite{cg:eigenwert}, we conclude:
\bt
There exist countably many eigenfunctions $(u_i)$ with eigenvalues $(\lam_i)$ such that the eigenvalues can be ordered
\begin{equation}
\lam_i\le \lam_{i+1}
\end{equation}
satisfying
\begin{equation}
\lim_{i\ra\un}\lam_i=\un,
\end{equation}
and such that their multiplicity is $1$. The eigenfunctions are smooth and they are complete in $\mc H_1$ as well as $L^2(\R[]_+)$.
\et

The multiplicity property is due to the fact that the differential equation is a second order ODE and that the eigenfunctions vanish in $r=0$.

\cvm
Next we consider the constrained eigenvalue problem for $H_2$. Let $\lam$ be an arbitrary eigenvalue for $H_1$, then we look at the implicit eigenvalue problem 
\begin{equation}
H_2\h-\bar\mu y^2\h-\lam \h=0,
\end{equation}
which is equivalent to
\begin{equation}
H_2\h-\lam\h=\bar\mu y^2\h,
\end{equation}
where $\bar\mu$ is supposed to be the eigenvalue.

First we distinguish two bilinear forms
\begin{equation}
\begin{aligned}
B(\h,\chi)=\spd{(H_2-\lam)\h}\chi=\int_{\R[]}(c_1 \h' \chi'+ V\h\chi-\lam \h\chi),
\end{aligned}
\end{equation}
where $c_1>0$, $\h,\chi\in C^\un_c(\R[])$, and
\begin{equation}
K(\h,\chi)=\int_{\R[]}y^2\h\chi.
\end{equation}

Note that
\begin{equation}
V(y)=2\al_M(\tilde\ka y+y^2)^2,
\end{equation}
hence $V$ growth like $y^4$ and we infer that the form $K$ is compact relative to $B$ in the space
\begin{equation}
\mc H_2=\set{\h\in L^2(\R[])}{\norm\h^2_2=\int_{\R[]}(\abs{\h'}^2+y^4\h^2)<\un}.
\end{equation}

Arguing as before we conclude
\bt
The eigenvalue problem
\begin{equation}
B(\h,\chi)=\bar\mu K(\h,\chi)\qq\A\,\chi\in\mc H_2
\end{equation}
has countably many solutions $(\bar\m_i,\h_i)$ such that
\begin{equation}
\bar\mu_i\le\bar\mu_{i+1},
\end{equation}
\begin{equation}
\lim_{i\ra\un}\bar\mu_i=\un,
\end{equation}
with multiplicity $1$. The eigenfunctions are smooth, vanish at infinity, and are complete in $\mc H_2$ as well as $L^2(\R[])$.
\et

\section{The spectral resolution}\las{6}

Let $(\lam,u)$ \resp $(\bar\mu,\h)$ satisfy
\begin{equation}
H_1 u=\lam u,
\end{equation}
\resp
\begin{equation}
H_2\h-\lam\h=\bar\mu y^2\h,
\end{equation}
then
\begin{equation}
\psi=u\h
\end{equation}
solves
\begin{equation}\lae{6.4}
H_2\psi-H_1\psi=\bar\mu y^2\psi,
\end{equation}
i.e., $\psi$ is a solution of the Wheeler-DeWitt equation.

Moreover,
\begin{equation}\lae{6.5}
\dot\psi=\dot u\h\q\wed\q \psi'=u\h',
\end{equation}
hence,
\begin{equation}\lae{6.6}
\int_{\R[]_+\times\R[]}\abs{D\psi}^2=\int_{\R[]_+}\abs{\dot u}^2\int_{\R[]}\abs\h^2 + \int_{\R[]_+}\abs u^2\int_{\R[]}\abs{\h'}^2,
\end{equation}
and similarly,
\begin{equation}\lae{6.7}
\int_{\R[]_+\times\R[]}\abs\psi^2y^p=\int_{\R[]_+}\abs u^2\int_{\R[]}\abs\h^2y^p,
\end{equation}
for $p=2,4$, as well as
\begin{equation}
\int_{\R[]_+\times\R[]}\abs\psi^2r^q=\int_{\R[]_+}\abs u^2r^q\int_{\R[]}\abs\h^2,
\end{equation}
where $q$ is the exponent in \fre{5.70}.

Thus, $\psi$ has bounded norm
\begin{equation}
\norm\psi^2=\int_{\R[]_+\times\R[]}\abs{D\psi}^2+\int_{\R[]_+\times\R[]}\abs\psi^2(r^q+y^4).
\end{equation}

Let $\mc H$ be the completion of $C^\un_c(\R[]_+\times\R[])$ with respect to this norm, then $\mc H$ can be viewed as  a dense subspace of  
\begin{equation}
\mc H_0=\set{\psi\in L^2(\R[]_+\times\R[])}{\norm\psi_0^2=\int_{\R[]_+\times\R[]}\abs\psi^2y^2},
\end{equation}
and the eigenfunctions $(\bar\mu_i,\psi_i)$ of \re{6.4} are complete in $\mc H$ as well as $\mc H_0$, where we note that the eigenfunctions are products
\begin{equation}
u_j\h_k.
\end{equation}

The claim that the eigenfunctions are complete needs some verification.
\bl\lal{5.10}
The eigenfunctions $\psi_i$ are complete in $\mc H$ as well as in $\mc H_0$.
\el

\bp
It suffices to prove the density in $\mc H$. The eigenfunctions are certainly complete in the closure of $C^\un_c(\R[]_+)\otimes C^\un_c(\R[])$ in $\mc H$, in view of \re{6.5} and \re{6.6}, but $C^\un_c(\R[]_+)\otimes C^\un_c(\R[])$ is dense in $\mc H$ as can be easily proved with the help of the Weierstra{\ss} approximation theorem.  
\ep

From now on we shall assume that the functions are complex valued. Denote by $A$ the symmetric operator
\begin{equation}
A=H_2-H_1
\end{equation}
with domain $D(A)\su\mc H_0$ equal to the subspace generated by its eigenfunctions. Define
\begin{equation}
\spd\psi\chi_0=\int_{\R[]_+\times\R[]}\psi\bar\chi y^2
\end{equation}
and
\begin{equation}
\spd{A\psi}\chi=\int_{\R[]_+\times\R[]}A\psi\bar\chi.
\end{equation}
Let $\psi$ be an eigenfunction, then
\begin{equation}\lae{6.14}
\spd{A\psi}\chi=\bar\mu\spd\psi\chi_0\qq\A\,\chi\in D(A),
\end{equation}
hence
\begin{equation}
\abs{\spd{A\psi}\chi}\le c_\psi \norm\chi_0,
\end{equation}
and this estimate is also valid for all $\psi\in D(A)$. Thus, the functional
\begin{equation}
\spd{A\psi}{\,\cdot\,}
\end{equation}
is continuous and anti-linear, and we deduce that there exists $\tilde\psi\in \mc H_0$ such that
\begin{equation}
\spd{A\psi}\chi=\spd{\tilde\psi}\chi_0\qq\A\,\chi\in D(A).
\end{equation}

\bl
The relation
\begin{equation}
\psi\ra\tilde\psi,\qq \psi\in D(A),
\end{equation}
defines a linear symmetric operator $\tilde A$ in $\mc H_0$ with domain $D(\tilde A)=D(A)$ satisfying
\begin{equation}\lae{6.19}
\tilde A\psi=\bar\mu\psi
\end{equation}
whenever $(\bar\mu,\psi)$ satisfies \re{6.4}.
\el
\bp
(i) $\tilde A$ is a mapping, since $D(A)$ is dense, and obviously linear.

\cvm
(ii) $\tilde A$ is symmetric, since for $\psi,\chi\in D(A)$
\begin{equation}
\spd{\tilde A\psi}\chi_0=\spd{A\psi}\chi=\spd\psi{A\chi}=\spd\psi{\tilde A\chi}_0.
\end{equation}

(iii) The relation \re{6.19} follows from \re{6.14}. 
\ep

\bl
$\tilde A$ is essentially self-adjoint in $\mc H_0$.
\el
\bp
It suffices to prove that $R(\tilde A\pm i)$ is dense, which is evidently the case, since the eigenfunctions belong to $R(\tilde A\pm i)$.
\ep

Let $H$ be the closure of $\tilde A$, then $H$ is self-adjoint and the spectral resolution for the Wheeler-DeWitt equation accomplished.

The Schr\"odinger equation for $H$ offers a dynamical development of the system provided the initial value is a finite superposition of eigenfunctions, \cf the remarks at the end of \cite[Section 8]{cg:qfriedman}.



\providecommand{\bysame}{\leavevmode\hbox to3em{\hrulefill}\thinspace}
\providecommand{\href}[2]{#2}



\end{document}